\documentclass[pra,twocolumn,amsmath,amssymb]{revtex4-1} %

\usepackage{epsfig,amsmath}
\usepackage{subfigure}
\usepackage{graphicx}% Include figure files
\usepackage{dcolumn}% Align table columns on decimal point
\usepackage{stmaryrd}
\usepackage{mathrsfs}
\usepackage{pifont}
\usepackage{amsthm}
\usepackage{amssymb}
\usepackage{bm}
\usepackage{latexsym}
\usepackage{hyperref}
\usepackage{color}

\newcommand{\beq}{\begin{equation}}
\newcommand{\eeq}{\end{equation}}
\newcommand{\beqa}{\begin{eqnarray}}
\newcommand{\eeqa}{\end{eqnarray}}
\newcommand{\la}{\langle}
\newcommand{\ra}{\rangle}

\newcommand{\ga}{\gamma}
\newcommand{\Ga}{\Gamma}

\newcommand{\da}{\dagger}
\newcommand{\De}{\Delta}

\newcommand{\om}{\omega}

\newcommand{\de}{\delta}

\newcommand{\non}{\nonumber}
\newcommand{\pa}{\partial}

\def\pra#1{{ Phys.\ Rev. A\/} {\bf#1}}

\def\prl#1{{ Phys.\ Rev.\ Lett.} {\bf#1}}

\def\pla#1{{ Phys.\ Lett. A\/} {\bf#1}}

\def\nat#1{{ Nature} {\bf#1}}

\begin{document}

\title{Non-Perturbative Quantum Dynamical Decoupling}

\author{Jun Jing$^{1,4}$\footnote{Email address: Jun.Jing@stevens.edu}, Lian-Ao Wu$^{2}$\footnote{Email address: lianao_wu@ehu.es}, J. Q. You$^{3}$\footnote{Email address: jqyou@fudan.edu.cn}, Ting Yu$^{1}$\footnote{Email address: Ting.Yu@stevens.edu}}

\affiliation{$^{1}$Center for Controlled Quantum Systems and Department of Physics and Engineering Physics, Stevens Institute of Technology, Hoboken, New Jersey 07030, USA\\ $^{2}$Ikerbasque, Basque Foundation for Science, 48011 Bilbao and Department of Theoretical Physics and History of Science, The Basque Country University (EHU/UPV), PO Box 644, 48080 Bilbao, Spain \\ $^{3}$Department of Physics and State Key Laboratory of Surface Physics, Fudan University, Shanghai 200433, China\\ $^{4}$Department of Physics, Shanghai University, Shanghai 200444, China}

\date{\today}

%%%%%%%%%%%%%%%%%%%%
\begin{abstract}
Current dynamical control based on the bang-bang control mechanism involving various types of pulse sequences is essentially a perturbative theory. This paper presents a non-perturbative dynamical control approach based on the exact stochastic Schr\"odinger equation. We report our findings on the pulse parameter regions in which the effective dynamical control can be exercised. The onset of the effective control zones reflects  the non-perturbative feature of our approach. The non-perturbative methods offer possible new implementations when several different parameter regions are available.
\end{abstract}


\maketitle

\section{Introduction}
%%%%%%%%%%%%%%%%%%%%%
One of the difficult issues in modern quantum science and technology is to overcome decoherence, i.e., the loss of coherence in a quantum system due to the interaction with its environment. Although theoretical and experimental studies done in this field have intensified over recent years, the results are still far from satisfactory. A variety of theoretical proposals for combating the deleterious environmental noise have been proposed \cite{Wiseman1,Wiseman2,Kurizki1,Law,Viola1,Tarn,Sanders,Kurizki2,Joynt}. Among them, dynamical decoupling or dynamical control, developed from the Bang-Bang method, of system-environment interactions by external fields is widely discussed due to its simplicity and accessibility to theoretical and experimental investigations \cite{Viola2,Uhrig1,Uhrig2,Lidar}. Notably,  the current theoretical formalism for the dynamical decoupling has been widely employed in combating decoherence and protecting information leakage, but most analyses and numerical simulations are based on idealized pulses and the Trotter product formula \cite{Nielsen}, where it is assumed that these idealized pulses are so fast and strong that the system-bath Hamiltonian can be effectively turned off when the pulses are applied to the system.

Ideal pulses by design can optimally protect the interested quantum state, but they are difficult to be implemented experimentally for most of practical setups. More specifically, if the control Hamiltonian is described by $H_{\rm c}=J\sigma_z$ and the original total Hamiltonian for system and environment is $H_{\rm tot}$, then the evolution operator combining the control mechanism becomes $U(\de)=\exp(-iH_{\rm tot}\de-iJ\sigma_z\de)\approx-i\sigma_z$ when $J\de=\pi/2$ in a short time $\delta$. Under this condition, the strength $J$ goes to infinity when $\de$ goes to zero. This amounts to taking the zero-order perturbation for an effective Hamiltonian $H_{\rm eff}\de =\pi/2\sigma_z+H_{\rm tot}\de$ (setting $\hbar =1$) with the small perturbation parameter $\de$. Essentially, all the current versions of dynamical control theories are based on a similar perturbation. Although the higher order perturbations have been considered in recent publications (e.g., see \cite{Uhrig2}),  yet without comparison with exact (numerical or analytical) solutions, it is difficult to know the validity range of the zeroth-order perturbation or higher order perturbation theory for a physical model with realistic system-bath interactions. Therefore, it is desirable to extend the standard perturbative dynamical coupling theory to a more general domain where non-ideal pulses can be employed in the control process.

The paper is organized as following.  Sec.~\ref{theory} introduces the theoretical framework of the non-perturbative dynamical control based on the non-Markovian quantum state diffusion (QSD) equation. It is put forward by controlling the fidelity of a single qubit system and the entanglement of a two-qubit system in a common non-Markovian dissipation model. In Sec. \ref{ent}, we present the open system control result with the effective zone of the control parameters that is the most distinguished feature given by the non-perturbative control theory. In Sec. \ref{dis}, we discuss the effect of different combination of control parameters and the memory time of the environment on the control efficiency. And finally we conclude all of the paper in Sec.~\ref{con}.

\section{Non-perturbative Dynamical Control} \label{theory}

Based on the exact quantum state diffusion equation \cite{Diosi1,Diosi2,Strunz}, we will perform theoretical analysis of the dynamical control theory beyond the perturbative dynamical control theories. Our approach employs the non-idealized pulses with finite widths $\Delta$, periods $\tau$ and finite pulse strengths $\Psi/\De$ as characteristic parameters. Interestingly,  we have found some efficient parameter regions where the quantum dynamics can be effectively controlled. It should be emphasized that the existence of the large parameter regions is purely due to the non-perturbative features of our approach. Moreover, we show that the characteristic quantity for quantum control is the ratio between the pulse period $\tau$ and the duration width $\Delta$. As such, the other pulse parameters, such as pulse strength, do not play a very essential role in this scheme \cite{Wu09}. Our results show that the parameters for idealized (or quasi-idealized) fast-strong pulses control ($\Delta$ is effectively taken as zeros) is a limiting case corresponding to a very small portion of the parameter region in which efficient quantum control can be realized. It will be shown that the control parameter regions for the non-perturbative approach impose much less severe constraints in implementing the dynamical control experimentally. In addition, we find that the non-Markovian environment is crucial for the dynamical control of quantum coherence.

\subsection{Non-Markovian quantum state diffusion equation}

Consider an open quantum system coupled to a general bosonic environment with the total Hamiltonian,
\begin{equation}
H_{\rm tot}=H_{\rm sys}+\hbar \sum_k(g_k^*La^\da_k+g_kL^\da a_k)+\sum_k\hbar\om_k a^\da_ka_k,
\end{equation}
where $H_{\rm sys}$, $L$ and $a_k (a_k^\dag)$ are the system Hamiltonian, coupling operator, and annihilation (creation) operator for mode $k$ of the environment, respectively. If the state vector $|\Psi(t)\rangle$  is used to denote the solution to the Schr\"{o}dinger equation with the total Hamiltonian in an interaction picture with respect to $H_{\rm env}=\sum_k\hbar\om_k a^\da_ka_k$, and define $|\psi_t(z^*)\rangle=\langle z| \Psi(t)\rangle$ where $|z\rangle=|z_1\rangle |z_2\rangle |z_3\rangle ....$ stands for the tensor product of the (Bargmann) coherent states for the environment modes, the dynamic equation for $|\psi_t(z^*)\rangle$ may be derived without any approximations, called non-Markovian quantum state diffusion equation \cite{Diosi1,Diosi2,YDGS99,Wiseman3,Strunz-Yu2004,Yu2004,Jing}. Explicitly, $|\psi_t(z^*)\rangle$ (a shorthand notation $\psi_t$) is governed by the following stochastic differential equation (setting $\hbar=1$) :
\begin{eqnarray}\non
 \partial_t\psi_t &=&[-i H_{\rm sys} + Lz_t^*- L^\da\bar{O}(t,z^*)]\psi_t
\equiv -i H_{\rm eff}\psi_t, \\\label{QSD}
z_t^*&=&-i\sum_kg_k^*z_k^*e^{i\om_kt},
\end{eqnarray}
where $H_{\rm eff}= H_{\rm sys} + iLz_t^*- iL^\da\bar{O}(t,z^*)$ is called an effective Hamiltonian, and  $z^*_t$ is a complex Gaussian process \cite{Zoller} describing the environmental influence, satisfying $M[z_t^*]=M[z_t^*z_s^*]=0$ and $M[z_tz_s^*] =\alpha(t,s)=\sum_k |g_k|^2 e^{-i\omega_k (t-s)}$. Note that $M$ denotes the statistical mean over the noise $z_t^*$ and $\alpha(t,s)$ is the environmental correlation function at zero temperature. For the standard Markov limit, one has $\alpha(t,s)=\Gamma \delta(t-s)$, i.e., the noise $z_t^*$ becomes a white noise and Eq.~(\ref{QSD}) reduces to a Markov stochastic Schr\"odinger equation \cite{Gisin1992}, which is equivalent to the corresponding Lindblad master equation or the quantum jump simulation \cite{Breuer}. Note that
 \beq\label{Oop}
 \bar{O}(t,z^*)\psi_t\equiv\int_0^t ds\alpha(t,s)O(t,s,z^*)\psi_t=\int_0^tds\alpha(t,s)\frac{\delta\psi_t}{\delta z_s^*}
 \eeq
is a polynomial function of operators acting on the system Hilbert space, which is determined by the following equation \cite{Diosi1,Diosi2},
\begin{equation}\label{cc}
\pa_tO(t,s,z^*)=[-iH_{\rm eff}, O(t,s,z^*)]-L^\da\frac{\delta\bar{O}(t,z^*)}{\delta z_s^*},
\end{equation}
with the initial condition $O(t=s,s,z^*)=L$. O-operator serves as an ``ansatz" to the functional derivative in Eq.~(\ref{Oop}), which represents the system response to the environmental noise and transforms the QSD equation into a convolutionless form. For a given model, once the exact O-operator is determined from Eq.~(\ref{cc}), then we obtain a time-local exact QSD equation, which is irrespective to the coupling strength between the system and environment and the environmental correlation function. Note that the reduced density matrix of the system of interest can be recovered by taking statistical mean over the noise,
\beq
\rho_t=M[|\psi_t(z^*)\ra\la\psi_t(z^*)|].
\eeq

The non-Markovian QSD equation is valid for the bosonic environment with an arbitrary correlation function \cite{Diosi2}. For simplicity, the non-Markovian environmental noise throughout the paper is described by the well-known Ornstein-Uhlenbek type noise with the correlation function $\alpha(t,s)=\frac{\Ga\gamma}{2}e^{-\gamma|t-s|}$, where $\Ga$ is the coupling strength between system and environment and  $1/\gamma$ characterizes the memory time of the non-Markovian environment. $\ga$ is non-tunable for a given environment. Notably, the correlation function has a well-defined Markov limit when $\gamma\rightarrow\infty$.

\subsection{One-qubit Case}

Here we first consider a single-qubit in a dissipative environment: $H_{\rm sys}=\frac{E(t)}{2}\sigma_z$ and $L=\sigma_-$, where $E(t)$ is typically a time-dependent function when a control mechanism is introduced. For this model, in the basis $\{|0\ra,|1\ra\}$, the effective Hamiltonian becomes
\begin{equation}
H_{\rm eff}=\left(\begin{array}{cc}
      -E/2 & iz_t^* \\
      0 & E/2-iF
    \end{array}\right),
\end{equation}
where $F(t)$ is the coefficient function contained in the exact O-operator $\bar{O}(t,z^*)=F(t)L$ (\cite{Diosi2}) and satisfies
\begin{equation}\label{Ft}
\partial_tF(t)=\Ga\gamma/2+(-\gamma+iE)F+F^2,
\end{equation}
and $F(0)=0$. To measure the survival probability of the initial state $\psi_0$ under the dissipative noise, the fidelity is defined by
\begin{equation}
\mathcal{F}(t)\equiv\la\psi_0|\rho_t|\psi_0\ra
=M[\la\psi_0|\psi_t(z^*)\ra\la\psi_t(z^*)|\psi_0\ra].
\end{equation}
When the initial state is chosen as $|\psi_0\ra=\mu|1\ra+\nu|0\ra$, $|\mu|^2+|\nu|^2=1$, in the rotating frame of $H_{\rm sys}$, the fidelity is easily obtained as
\begin{eqnarray}\non
\mathcal{F}(t)&=&1-|\mu|^2-(|\mu|^2-2|\mu|^4)
e^{-2\int_0^tds\mathcal{R}[F(s)]}
\\ \label{Fi} &+&2|\mu|^2(1-|\mu|^2)\mathcal{R}[e^{-\int_0^tdsF(s)}]
\end{eqnarray}
where ${\mathcal R}[\cdot]$ stands for the real part of a complex function. Under the condition that $\int_0^tdsF(s)\rightarrow0$ (it can be realized by applying appropriate pulses as to be seen later), $\mathcal{F}(t)$ approaches unity, which is independent on the initial state.

\subsection{Two-qubit Case}

Next we consider a two-qubit system embedded in a collective dissipative environment with the following Hamiltonian,
\beq
H_{\rm sys}=\frac{E(t)}{2}(\sigma_z^A+\sigma_z^B), \quad  L=\sigma_-^A+\sigma_-^B.
\eeq
This model corresponds to the collective error considered in quantum information.

For this two-qubit system, the exact O-operator contained in the stochastic Schr\"{o}dinger equation (\ref{QSD}) can be explicitly derived by Eq.~(\ref{cc}) \cite{Zhao}:
\begin{equation}
\bar{O}(t,z^*)=F_1(t)O_1+F_2(t)O_2+iU_zO_3,
\end{equation}
where $O_1=L$, $O_2=\sigma^A_z\sigma^B_-+\sigma^A_-\sigma^B_z$, $O_3=\sigma^A_-\sigma^B_-$, and $U_z\equiv\int_0^tdsU(t,s)z_s^*$. The determination of these coefficient functions could be found in the following differential equation group:
\begin{eqnarray}\non
\partial_tF_1(t)&=&\Ga\gamma/2+(-\gamma+iE)F_1+F_1^2+3F_2^2
-i\bar{U}/2, \\ \non
\partial_tF_2(t)&=&(-\gamma+iE)F_2-F_1^2+4F_1F_2+F_2^2-i\bar{U}/2, \\
\label{FFU} \partial_t\bar{U}(t)&=&-2i\gamma
F_2+(-2\gamma+2iE)\bar{U}+4F_1\bar{U},
\end{eqnarray}
with the notation $\bar{U}(t)\equiv\int_0^tds\alpha(t,s)U(t,s)$. The boundary conditions are given by $F_1(0)=F_2(0)=\bar{U}(0)=0$ and $U(t,t)=-4iF_2(t)$.

For this model, it is convenient to choose a new basis $\{|s\ra,|a\ra,|11\ra,|00\ra\}$ where $|s\ra=(1/\sqrt{2})(|10\ra+|01\ra)$ and $|a\ra=(1/\sqrt{2})(|10\ra-|01\ra)$. In the new basis, the stochastic Hamiltonian can be rewritten as,
\begin{equation}\label{Heff2}
H_{\rm eff}=\left(\begin{array}{cccc}
 -2if & 0  & \sqrt{2}(iz_t^*+U_z) & 0 \\
 0 & 0 & 0 & 0 \\
 0 & 0 & E-2i(F_1+F_2)  & 0 \\
 \sqrt{2}iz_t^* & 0 & 0 & -E
    \end{array}\right),
\end{equation}
where $f(t)\equiv F_1(t)-F_2(t)$.  One can check that
\begin{equation}\label{ft}
\partial_tf(t)=\Ga\gamma/2+(-\gamma+iE)f+2f^2,
\end{equation}
and $f(0)=0$ from Eq.~(\ref{FFU}). An initial state (pure state) belongs to the subspace spanned by the first two basis vectors $|s\ra$ and $|a\ra$ may be written as
\beq
|\psi_0\ra=P_s(0)|s\ra+P_a(0)|a\ra+q_{11}(0)|11\ra+q_{00}(0)|00\ra
\eeq
 with $|P_s(0)|^2+|P_a(0)|^2=1$ and $q_{11}(0)=q_{00}(0)=0$. From Eqs.~(\ref{QSD}) and (\ref{Heff2}), we have
 \beq
 |\psi_t\ra=P_s(t)|s\ra+P_a(t)|a\ra+q_{00}(t)|00\ra,
 \eeq
where  $P_s(t)=e^{-2\int_0^tdsf(s)}P_s(0)$ and $P_a(t)=P_a(0)$. Note $|a\ra$ is found to be in the decoherence free subspace \cite{Lidar2} of the effective Hamiltonian (\ref{Heff2}) due to its vanishing elements insider the $2$nd dimension. The Wootters concurrence \cite{Wootters} for each quantum trajectory is given by
\begin{equation}\label{Cpsi}
C(\psi_t)=|e^{-4\int_0^tdsf(s)}P_s^2(0)-P_a^2(0)|.
\end{equation}
In fact, Eqs.~(\ref{ft}) and ~(\ref{Cpsi}) are noise-independent. Statistical average yields $M[C(\psi_t)]=C(\psi_t) $ by Novikov theorem \cite{YDGS99}. Therefore, we have
 \beq
 C(\rho_t)=|\rho_{ss}-\rho_{aa}-\rho_{sa}+\rho_{as}|,
 \eeq
where $\rho_{ss}=M[|P_s(t)|^2]$, $\rho_{aa}=M[|P_a(0)|^2]$, and $\rho_{sa}=\rho_{as}^*=M[P_s^*(t)P_a(0)]$. Straightforward calculation shows that $C(\rho_t)=M[C(\psi_t)]$. Moreover, $C(\rho_t)$ does not depend on $\rho_{00}$ and any coherence terms involving $|00\ra$. Thus for any initial states with $q_{11}(0)=0$, Eq.~(\ref{Cpsi}) is valid and is always equal to $C(\rho_t)$.

Interestingly, we can also get an explicit expression for the fidelity of the state:
\begin{eqnarray}\label{Fi2}
{\mathcal F}(t)=|e^{-2\int_0^tdsf^*(s)}|P_s^2(0)|+|P_a^2(0)||^2.
\end{eqnarray}
Note that for both concurrence (\ref{Cpsi}) and fidelity (\ref{Fi2}), the two physical quantities will be frozen to their original values when $\int_0^tdsf(s)\rightarrow0$. It is easy to see that the results are independent on the initial states in the subspace of interest.

It should be noted that if $|P_s(0)|=1$ and $P_a(0)=0$, i.e. $|\psi_0\ra=|s\ra$,
\beq
C(\psi_t)={\mathcal F}(t)=e^{-4\int_0^tds{\mathcal R}[f(s)]}.
\eeq
In this situation, the fidelity can be used as a reliable measure for entanglement.

\section{Control of Open Systems} \label{ent}

The exact quantum state diffusion equation (\ref{QSD}) provides a very useful theoretical framework for the non-perturbative dynamical decoupling approach. The external control on the system of interest is introduced through the system energy shift $E(t)=\om+c(t)$, where $\om$ is the bare frequency and $c(t)$ is a time-dependent control function. In this way, the external pulse control, as well as all of the characteristic parameters,  has been naturally combined into the system Hamiltonian. Once we find the exact QSD equation, the controlled dynamics can be formally treated as a quantum open system problem. This allows us to simulate non-perturbative dynamical decoupling dynamics with a wider range of control parameters without any approximation.

In principle, the control $c(t)$ can be an arbitrary function of time that can be realized by an external field applied to the system, including rectangular pulse sequences with arbitrary strength and frequency. The zeroth-order perturbation or idealized pulse approximation is shown to suppress the information leakage rate of subspace to the environment or the rest part of the system if the frequency of the pulse is sufficiently high. The focus of this paper is to explore the non-perturbative regimes beyond the conventional perturbation and the idealized pulses.

The control parameters considered in this paper consist of the period of the rectangular pulse $\tau$, the duration $\De$ (or the dark duration $\tau-\De$) and the strength $\Psi/\De$, respectively, i.e. $c(t)=\Psi/\De$ for regions $n\tau-\De<t\leqslant n\tau$, $n\geqslant1$ integral, otherwise $c(t)=0$. The pulse frequency is determined by the dimensionless proportion parameter $\tau/\De$. With the same $\De$, the smaller $\tau/\De$, the more frequent pulses.

\begin{figure}[htbp]
\centering
\includegraphics[width=3.2in]{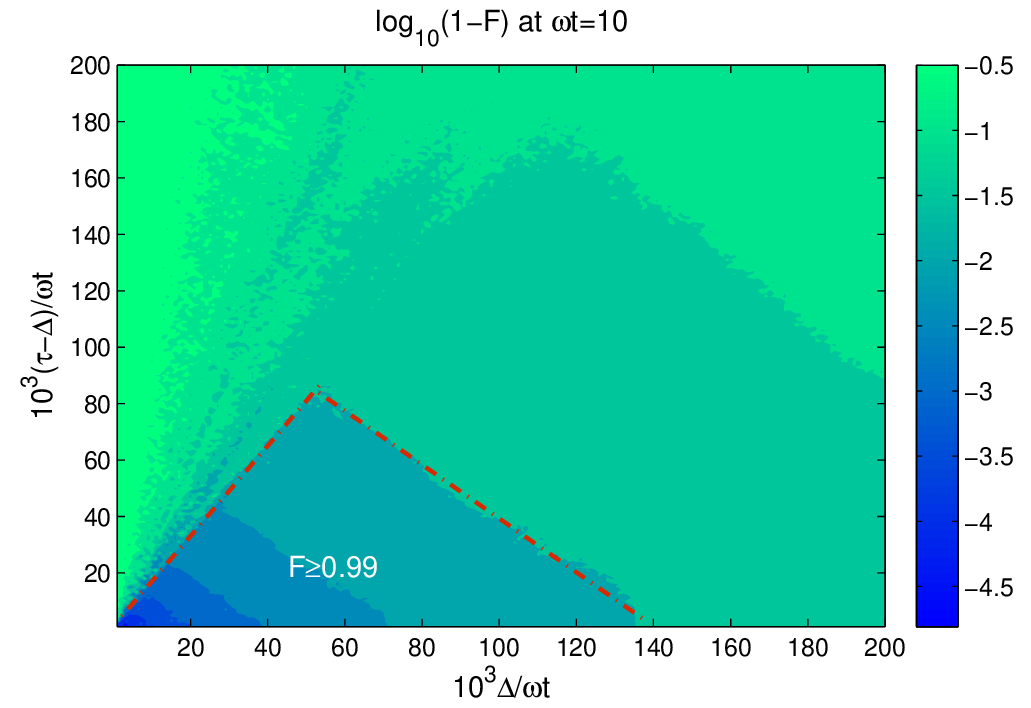}
\caption{$\log_{10}(1-\mathcal{F})$ at the time point $\omega t=10$ of the single-qubit system under control. It is initially prepared in $|\psi_0\ra=(1/\sqrt{2})(|1\ra+|0\ra)$. Here we choose $\Psi=\om$, $\Ga=1$, and $\ga=0.5$.}\label{one}
\end{figure}

In Fig.~\ref{one}, we plot the $\De-\tau$ diagram of the fidelity control a single qubit at $\om t=10$. By definition, $\tau$ is always larger than $\De$ (otherwise if $\tau=\De$, then the pulse sequence becomes a constant field), we numerically compute the fidelity (\ref{Fi}) with different parameter ratios of $\De$ and $\tau-\De$. Notably, the ideal pulse with infinitesimal width occupies only the left-down corner of this diagram. In this figure, the whole parameter space is clearly divided into several regions according to the value of $\log_{10}(1-\mathcal{F})$ at the given time $\omega t=10$. Roughly from the left-down corner to right upper corner, we could see continuous enlarged triangular zones, whose color suggest the fidelity with $\mathcal{F}\geq0.9999$, $\mathcal{F}\geq0.999$, $\mathcal{F}\geq0.99$ and $\mathcal{F}\geq0.9$, respectively. The results clearly show there exists a very large parameter space that allows an efficient controllability of fidelity. We can see that the efficient control region of ${\mathcal F}\geq0.99$ occupy a very large parameter space beyond the ideal pulse sequence. The duration time of each pulse could be even chosen as large as $0.135\om t$.

\begin{figure}[htbp]
\centering
\subfigure{\label{two:dp}
\includegraphics[width=3.2in]{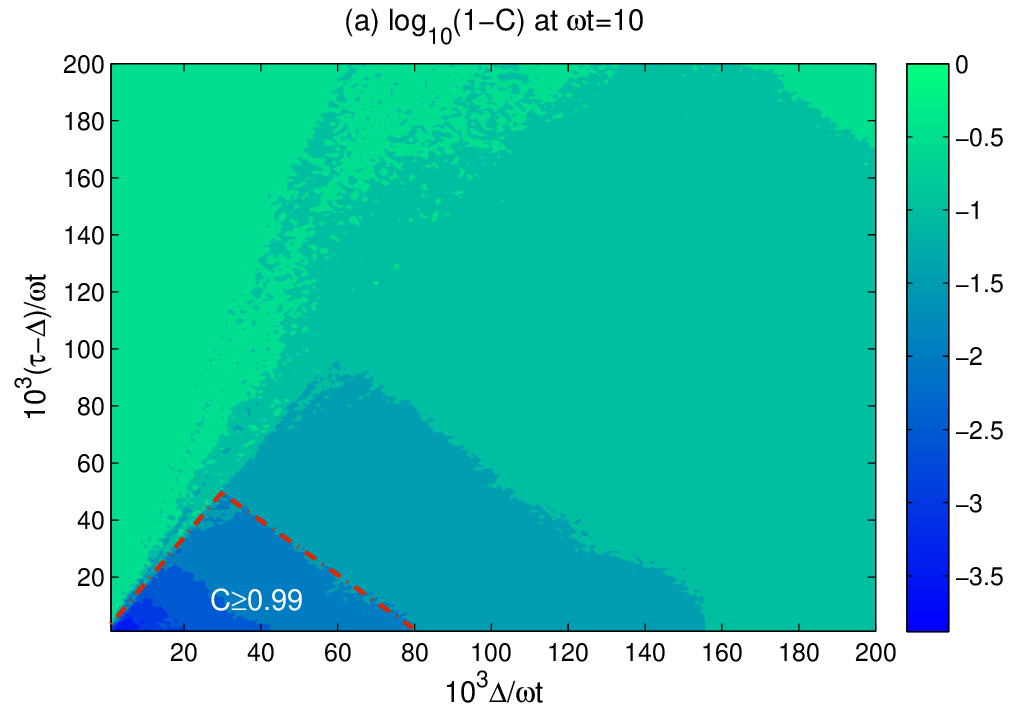}}
\subfigure{\label{two:str}
\includegraphics[width=3.2in]{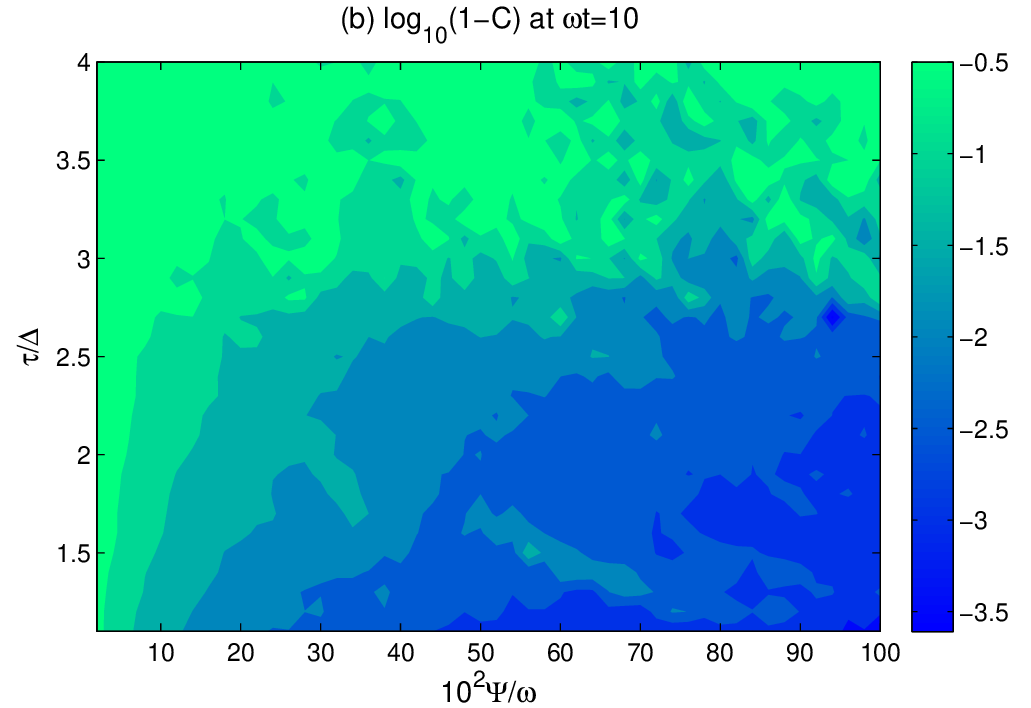}}
\caption{$\log_{10}(1-C)$ at the time point $\omega t=10$ of a two-qubit system under control. It is initially prepared in $|\psi_0\ra=|s\ra$. We choose $\Ga=1$ and $\ga=0.5$. (a) $\De-\tau$ diagram with $\Psi=\om$; (b) $\tau/\De-\Psi$ diagram with $\De=0.01\omega t$.}\label{two}
\end{figure}

In what follows, we will demonstrate that there exist some parameter regions that allow efficient control of quantum entanglement with a more practical pulse train. For simplicity, the initial state of the two-qubit system is taken as $|\psi_0\ra=|s\ra$. We should note the our non-perturbative quantum control is not limited to this particular state. In fact, our results can be extended to an arbitrary state. Without the external pulse control, the state will lose its entanglement eventually in time. Figure \ref{two} plots two parameter-diagrams of concurrence at the time point $\om t=10$.  Figure \ref{two}(a) describes the $\De-\tau$ diagram. The control efficiency regions denoting $C\geq0.999$, $C \geq 0.99$ and $C \geq 0.9$, also take expanding triangular configurations. Here the boundary of the width of pulse duration where $C \geq 0.99$ could be relaxed up to about $0.08\om t$, although in this case the dark time is very limited. In the zone of $C\geq0.99$, $\tau/\De$ could be up to about $2.5$; otherwise, at this very moment, the system state will fall into other zones with lower concurrence. Figure \ref{two}(b) is the diagram of $\tau/\De-\Psi$, the ratio of duration time and period of the pulse v.s. the strength of the pulse. It clearly shows that another important desired factor of idealized-pulse, the infinite strong pulse, is not as so necessary as one would expect from a perturbative point view. Here we see again that the infinite strong approximation is only a small portion of the large permissible parameter region. To realize efficient control $C\simeq1$, the ratio of $\tau/\De$ must be lower than $3$. The requirement to pulse strength is also important to some extend since the ideal pulse is necessary to get $C \geq0.999$. The pulse strength cannot be too weak. As least when $\Psi<0.15\om$, it is impossible to keep the concurrence as high as $0.99$. Yet this control target is accessible with an optimized ratio of $\tau/\De$, even when the strength is relaxed to $0.2\om$.

\section{Discussion}\label{dis}

\begin{figure}[htbp]
\centering
\includegraphics[width=3.2in]{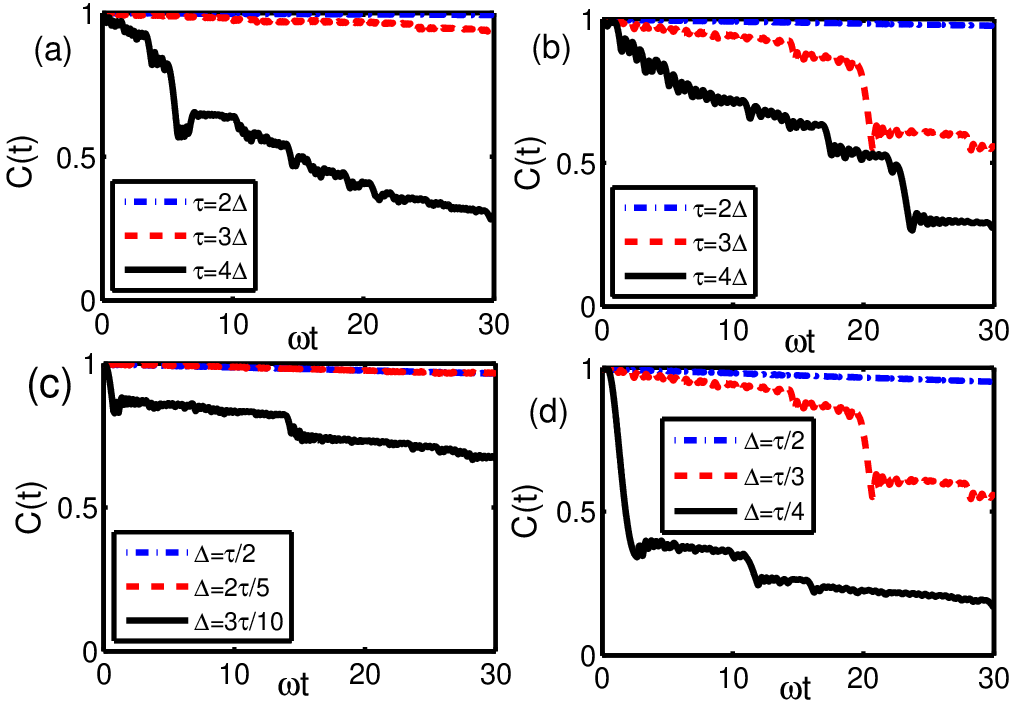}
\caption{The concurrence dynamics of a two-qubit system prepared in
$|\psi_0\ra=|s\ra$ with different $\De$'s and $\tau$'s. (a) $\Delta=10^{-2}\omega t$; (b) $\Delta=2\times10^{-2}\omega t$; (c) $\tau=5\times10^{-2}\omega t$; (d) $\tau=6\times10^{-2}\omega t$. We choose $\Psi=\om$, $\Ga=1$, and $\ga=0.5$.}\label{T11}
\end{figure}

\begin{figure}[htbp]
\centering
\includegraphics[width=3.2in]{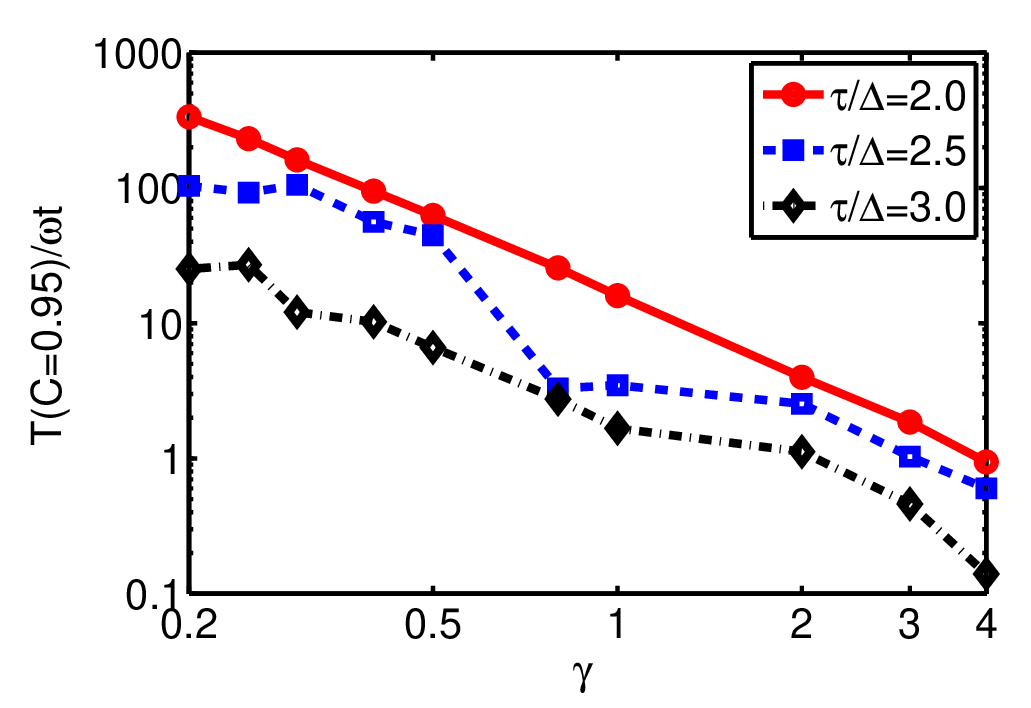}
\caption{The concurrence damping time $T(C=0.95)$ in the dimension of $\om t$ about a two-qubit system prepared in $|\psi_0\ra=|s\ra$ vs. $\gamma$ with different $\tau/\De$'s. We choose $\Delta=0.02\om t$, $\Ga=1$, and $\Psi=\om$.}\label{ga}
\end{figure}

The non-perturbative control may also be observed from a dynamical view point. Figure \ref{T11} shows the concurrence dynamics under different control parameters. In Fig.~\ref{T11}(a) and (b), we fix $\De$ and use different values of $\tau$ to show the effect of $\tau$; while in Fig.~\ref{T11}(c) and (d), we show the effect of different $\De$ with a fixed $\tau$. For a long time scale $\om t\leq30$, each value of $\De$ constrains the value of $\tau$ to yielding a good entanglement control [that is $C(t)$ is frozen as $C(0)=1$] and vice versa. We therefore conclude that as long as $\tau/\Delta$ is not too large, both entanglement and fidelity can be efficiently protected.

Note that Figs.~\ref{two} and \ref{T11} are plotted with $\ga=0.5$, which corresponds to a moderate non-Markovian regime. Figure \ref{ga} is used to show how the environmental memory time affects the non-perturbative  quantum control scheme. The presence of the non-zero environmental memory indicates a characteristic feature of a non-Markovian dynamics, which causes the  bidirectional flow of information between the open system and its environment.  More qualitatively, in so-called Markov limit, the memory time is zero, as such, the information of the system state (superposition, coherence, etc.) will quickly be lost irreversibly into the infinite environmental modes. The one-way information loss will not permit the external pulses to efficiently restore the initial open system state since the decoherence will quickly suppress the viable quantum coherence before the control mechanism becomes effective. Now we use $T$ to denote the time after which the concurrence decays from unity to $0.95$. Figure \ref{ga} is plotted in a logarithmic scale to yield a better view of the entanglement decay speed against $\ga$. We emphasize that the non-Markovian bath is essential for the effective quantum control as shown in Fig.~\ref{ga}.

\section{Conclusion}\label{con}

In conclusion, we have established a non-perturbative approach to dynamical control theory based on the non-Markovian quantum state diffusion approach. We have found the effective parameter regions of non-ideal pulse sequence where dynamical decoupling can be used to protect the fidelity and entanglement of arbitrary pure states in a system Hilbert space. As a limiting case, we see that the idealized pulse approximation belongs to a special part of the large permissible parameter region that is able to suppress the noise. This may suggest that one can go beyond the zeroth-order approximation and thus prompts more experimental implementations of the dynamical decoupling scheme in a wide spectrum of parameter ranges. Additionally, we have examined the environmental memory effect on the decoupling process.

Our non-perturbative approach can accommodate generic system-bath couplings, structured baths and varied numbers of environmental modes ranging from one mode, two modes to infinite number of modes. Our approach provides greater flexibility in the experimental implementation of the dynamical decoupling scheme since our scheme allows the existence of a much larger parameter region. In addition, the effectiveness of our approach is independent on the initial states of the system of interest. Our non-perturbative approach may open up a new avenue for more practical implementations of quantum control technique in quantum information processing.

\acknowledgments
We acknowledge grant support from the NSF PHY-0925174, AFOSR No.~FA9550-12-1-0001, the NSFC Nos. 91121015 and 11175110, an Ikerbasque Foundation Startup, the Basque Government (grant IT472-10) and the Spanish MICINN (Project No. FIS2009-12773-C02-02).

\end{document}